\documentclass[useAMS,usenatbib]{mn2e}
\usepackage{times}
\usepackage{graphics,epsfig}
\usepackage{graphicx}
\usepackage{amssymb}
\usepackage{multirow}

\title[A possible group of galaxies at {\boldmath $z$} = 0.069]{The curious case of J113924.74{\boldmath$+$}164144.0: a possible new group of galaxies at {\boldmath $z$} = 0.069} 
\author[Roy, Sengupta and Kantharia]{Nirupam Roy$^{1}$\thanks{E-mail: nroy@aoc.nrao.edu~(NR); chandra@caha.es~(CS); ngk@ncra.tifr.res.in~(NGK)}, Chandreyee Sengupta$^{2,3}$\footnotemark[1] and N. G. Kantharia$^{4}$\footnotemark[1]\\ 
       $^{1}$National Radio Astronomy Observatory, 1003 Lopezville Road, Socorro, NM 87801, USA\\
       $^{2}$Instituto de Astrofisica de Andalucia, C/ Camino Bajo de Huetor, 50, Granada 18008, Spain\\
       $^{3}$ Calar Alto Observatory, Centro Astronomico Hispano Aleman, C/ Jesus Durban Remon, 2-2, Almeria 04004, Spain\\
       $^{4}$National Centre for Radio Astrophysics, TIFR, Post Bag 3, Ganeshkhind, Pune 411 007, India}

\begin{document}
\date{Accepted yyyy month dd. Received yyyy month dd; in original form yyyy 
month dd}

\pagerange{\pageref{firstpage}--\pageref{lastpage}} \pubyear{2010}

\maketitle

\label{firstpage}

\begin{abstract}
J113924.74$+$164144.0 is an interesting galaxy at $z = 0.0693$, i.e. $D_L \sim 
305$ Mpc, with tidal-tail-like extended optical features on both sides. There 
are two neighbouring galaxies, a spiral galaxy J113922.85$+$164136.3 which has 
a strikingly similar `tidal' morphology, and a faint galaxy 
J113923.58$+$164129.9. We report H~{\sc i} 21 cm observations of this field to 
search for signatures of possible interaction. Narrow H~{\sc i} emission is 
detected from J113924.74$+$164144.0, but J113922.85$+$164136.3 shows no 
detectable emission. The total H~{\sc i} mass detected in 
J113924.74$+$164144.0 is $7.7\times10^9$ M$_\odot$. The H~{\sc i} emission 
from the galaxy is found to be extended and significantly offset from the 
optical position of the galaxy. We interpret this as signature of possible 
interaction with the neighbouring spiral galaxy. There is also a possible 
detection of H~{\sc i} emission from another nearby galaxy 
J113952.31$+$164531.8 at $z = 0.0680$ at a projected distance of $600$ kpc, 
and with a total H~{\sc i} mass of $5.3\times10^9$ M$_\odot$, suggesting that 
all these galaxies form a loose group at $z \sim 0.069$.
\end{abstract}


\section{Introduction}
\label{sec:int}

The 21 cm transition has long proved to be an extremely useful tool to probe 
the neutral hydrogen (H~{\sc i}) in both the local and the distant universe. 
In particular, the H~{\sc i} 21 cm emission has been used to study, in detail, 
properties of the interstellar medium (ISM) of the nearby galaxies in the 
local universe \citep[][and references therein]{br90,zw97,zw05}. Particularly 
for gas rich spiral galaxies (at low and moderate distances) where the 
H~{\sc i} disk is significantly bigger than the stellar disk, H~{\sc i} 
observations provide information not only on the dynamics of the particular 
galaxy of interest but also on the galaxy environment and on possible 
interactions of neighbouring galaxies 
\citep[e.g.,][]{ho90,ro90,ka05,ho07,se09}. Cosmic evolution of the properties 
of neutral gas in galaxies also have important implications on our 
understanding of star formation and galaxy evolution.

However, it should be noted here that, due to weakness of the emission signal, 
detecting H~{\sc i} emission from galaxies even at moderate redshift ($z 
> 0.05$) is a challenging task. Instead, H~{\sc i} absorption has been 
extensively used to probe ISM at high redshift 
\citep[e.g.,][]{ca96,br96,ka07,yo07} but only in the intervening gas along a 
narrow line of sight towards the background source and therefore can not be as 
informative as H~{\sc i} emission in the kinematics and the dynamics of the 
gas within galaxies.

\begin{figure*}
\begin{center}
\begin{tabular}{ll}
\multirow{3}{*}{\includegraphics[scale=0.67, angle=0.0]{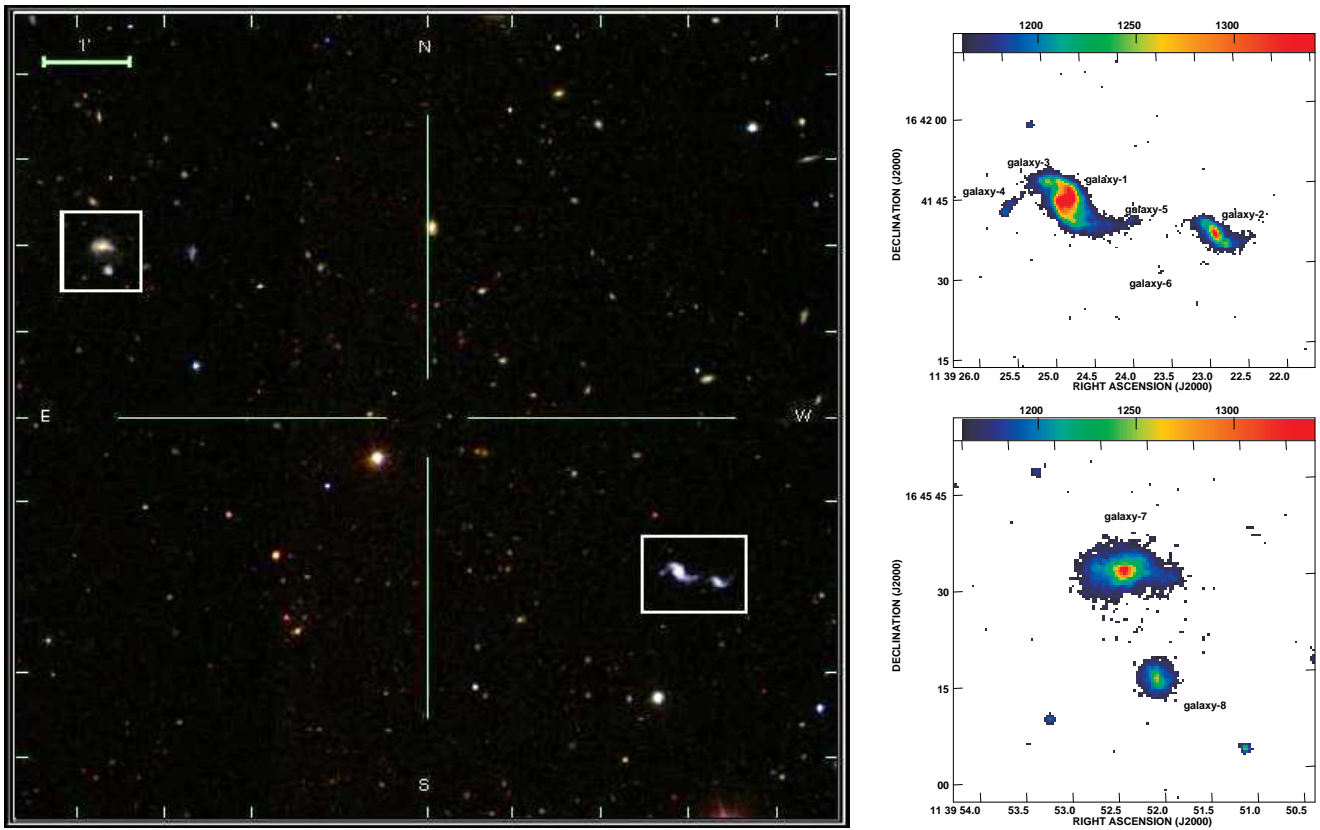}}&{\vtop{\vskip-1ex\hbox{{\includegraphics[scale=0.31, angle=0.0]{1134r.ps}}}}}\\
&{\includegraphics[scale=0.31, angle=0.0]{1135r.ps}}\\
\end{tabular}
\caption{\label{fig:test} {\it Left:} SDSS DR6 $r$-band image of the field ($9.5\arcmin \times 9.5\arcmin$) centred at RA 11:39:37 Dec.~+16:43:32 (J2000) showing two galaxy pairs. {\it Top right:} Part of the field showing details of the galaxy pair J113924.74$+$164144.0 and J113922.85$+$164136.3 (marked `galaxy-1' and `galaxy-2' respectively). Note the other four neighbour sources identified as separate galaxies in SDSS (`galaxy-3', `galaxy-4',`galaxy-5' and `galaxy-6'). {\it Bottom right:} Part of the field showing details of the galaxy pair J113952.31$+$164531.8 and J113952.02$+$164514.8 (marked `galaxy-7' and `galaxy-8' respectively). Note the striking morphological similarity of the two spirals (galaxy-1 and galaxy-2) and the disturbed morphology of galaxy-7. The GMRT radio observation field of view was $\sim 25\arcmin$ centred at galaxy-1. The optical properties for these sources are listed in Table~\ref{table:src1}.}
\end{center}
\end{figure*}

Recently, however, attempts to detect H~{\sc i} emission at high $z$ have been 
made with upgraded or new powerful receivers. Large H~{\sc i} surveys like 
the ongoing Arecibo Legacy Fast 
ALFA (ALFALFA) Survey, the H~{\sc i} Parkes All-Sky Survey (HIPASS), the 
H~{\sc i} Jodrell All-Sky Survey (HIJASS) and the Arecibo Galaxy Environment 
Survey (AGES) are targeting detailed study of H~{\sc i} only in nearby 
galaxies at $z \lesssim 0.06$ \citep{ba01,la03,gi05a,gi05b,au06}. In the last 
ten years, sensitive observations with very long integration times using radio 
telescopes like Arecibo, GMRT, VLA and WSRT have resulted in a few detection 
of H~{\sc i} emission from galaxies at $z \gtrsim 0.1$: a single galaxy in Abell 
2218 at $z = 0.18$ \citep{zw01}, one in Abell 2192 at $z = 0.19$ \citep{ve04}, 
19 galaxies in Abell 963 at $z= 0.21$ and 23 galaxies in Abell 2192 at $z = 
0.19$ \citep{ve07} and 20 optically selected galaxies from the Sloan Digital 
Sky Survey \citep{ca08} at redshifts between $z = 0.17 - 0.25$. Additionally, 
there have been measurements of neutral atomic hydrogen gas content from 
multiple galaxies using co-adding technique for galaxy clusters Abell 3128 at 
$z = 0.06$, Abell 2218 at $z = 0.18$, a sample of star-forming galaxies at 
$z= 0.24$ and Abell 370 at $z = 0.37$ \citep{zw00,ch01,la07,la09}. However, 
with hundreds of hours of observation time, all these are detection of mainly 
very gas rich systems and the number of H~{\sc i} emission measurements beyond 
the local universe is still small. Hence H~{\sc i} emission study of galaxies 
at $z > 0.05$ is challenging but important from many considerations including 
increasing the number of H~{\sc i} detection in individual galaxies at these 
redshifts and understanding the galaxy evolution.

Here we present H~{\sc i} observation and results of a particularly 
interesting galaxy J113924.74$+$164144.0 at $z = 0.0693$ and its neighbouring 
galaxy J113952.31$+$164531.8 at $z=0.0680$. The details of the field 
containing these galaxies are outlined in Section \S\ref{sec:src} while the 
observation and data analysis method is briefly described in Section 
\S\ref{sec:daa}. Section \S\ref{sec:rsl} contains the results, and we present 
conclusions in Section \S\ref{sec:con}.

\section{The field of J113924.74{\boldmath$+$}164144.0}
\label{sec:src}

\begin{table*}
 \caption{Details of the sources in the SDSS field with J113924.74$+$164144.0}
\begin{center}
 \begin{tabular}{lcccccccccc}
 \hline
ID & SDSS galaxy           & \multicolumn{2}{c}{Distance$^a$} & \multicolumn{5}{c}{Photometry} & Redshift & cz\\
   &                       & $\arcmin$& kpc$^b$ & u    & g     & r     & i     & z     & $z_{sp}$ & km~s$^{-1}$\\
\hline
galaxy-1  & J113924.74$+$164144.0 & ---   &  --   & 17.83 & 16.82 & 16.42 & 16.14 & 16.05 & 0.0693 & 20774\\
galaxy-2  & J113922.85$+$164136.3 & 0.471 & ~37 & 19.21 & 18.26 & 17.80 & 17.57 & 17.53 & --     & -- \\
galaxy-3$^c$  & J113925.00$+$164147.8 & 0.088 & ~~7 & 20.63 & 20.23 & 20.49 & 20.14 & 19.76 & -- & -- \\
galaxy-4$^c$  & J113925.55$+$164142.5 & 0.195 & ~15 & 20.41 & 20.02 & 20.03 & 19.67 & 19.89 & -- & -- \\
galaxy-5$^c$  & J113923.94$+$164139.5 & 0.206 & ~16 & 20.29 & 19.57 & 19.45 & 19.11 & 19.03 & -- & -- \\
galaxy-6  & J113923.58$+$164129.9 & 0.364 & ~28 & 22.19 & 22.50 & 21.81 & 21.59 & 23.38 & --     & -- \\
galaxy-7  & J113952.31$+$164531.8 & 7.615 & 595 & 19.35 & 17.60 & 16.82 & 16.42 & 16.08 & 0.0680 & 20386\\
galaxy-8  & J113952.02$+$164514.8 & 7.415 & 580 & 19.86 & 18.67 & 18.24 & 17.96 & 17.91 & --     & -- \\ 
\hline
\end{tabular}\\
\begin{flushleft}
$^a$ Distance from the spiral galaxy J113924.74$+$164144.0; ~~~$^b$ Projected separation at $z = 0.069$; ~~~$^c$ Most likely to be part of galaxy-1
\end{flushleft}
\end{center}
\label{table:src1}
\end{table*}

The spiral galaxy J113924.74$+$164144.0 ($m_r = 16.42$) at a spectroscopic 
redshift of $z_{sp}=0.069295 \pm 0.000077$ ($cz = 20774$ km~s$^{-1}$, 
corresponds to a look-back time of $\sim 1$ billion years) is found to be a 
very interesting object because of two main reasons. First of all, there are 
long, tidal-tail-like extended features on both sides of this galaxy (see 
Figure~\ref{fig:test} for $r$-band image). Though, these features bear strong 
resemblance to tidal structures, SDSS DR6 \citep{sd,dr6} has identified these 
structures as three separate faint 
galaxies: (1) J113925.00$+$164147.8 ($m_r = 20.49$; photometric redshifts from 
SDSS $z_{ph} \sim 0.99$, $0.46$ and $0.50$), (2) J113925.55$+$164142.5 ($m_r = 
20.03$; $z_{ph} \sim 1.00$, $0.47$ and $0.30$) and (3) J113923.94$+$164139.5 
($m_r = 19.45$; $z_{ph} \sim 0.04$, $0.13$ and $0.12$). Derived photometric redshift for these features quoted in SDSS has large error with a overlap around 0.07 indicating that they are probably at the same redshift. Figure~\ref{fig:test} 
shows the SDSS DR6 $r$-band image of the field containing this galaxy along 
with a $\sim 2 \times 2\arcmin$ image around this main galaxy (marked as 
`galaxy-1' in the figure) to show the details of these three neighbouring 
``galaxies'' (`galaxy-3', `galaxy-4' and `galaxy-5' in the figure). Other than 
these three galaxies, there is another faint ($m_r = 21.81$) galaxy 
(`galaxy-6' in the figure), J113923.58$+$164129.9 ($z_{ph} \sim 0.22$, $0.60$ 
and $0.58$), which is $0.364\arcmin$ away and a comparatively bright 
($m_r = 17.80$) spiral galaxy (`galaxy-2' in the figure), 
J113922.85$+$164136.3 ($z_{ph} \sim 0.00$, $0.05$ and $0.09$), 
$0.471\arcmin$ away. This corresponds to a projected separation of $\sim 37$ 
kpc at $z=0.069$. Both galaxy-1 and galaxy-2 seems to have open 
spiral arms and no significant bulge, and so are likely to be Sc or Sd type. 
The second interesting point is that this smaller spiral galaxy has an optical 
morphology with tidal-tail-like extended features very similar to that of the 
bigger spiral galaxy. Given that the errors in photometric redshift 
measurements are large due to various uncertainties, it is possible that all 
these galaxies are almost at the same redshift and interacting with each 
other. Apart from these, $\sim 7\arcmin$ away from the source ($\sim 600$ kpc 
at $z=0.069$) there are two galaxies, J113952.31$+$164531.8 ($m_r = 16.82$) 
with a spectroscopic redshift 
of $z_{sp}=0.068000 \pm 0.000090$ ($cz = 20386$ km~s$^{-1}$) and its close neighbour 
J113952.02$+$164514.8 ($m_r = 18.24$; $z_{ph} \sim 0.04$, $0.06$ and $0.07$). 
SDSS DR6 $r$-band image of the field containing these two galaxies (marked 
`galaxy-7' and `galaxy-8') is also shown in Figure~\ref{fig:test}. Clearly, 
the optical morphology of galaxy-7 is ``disturbed'' with tidal features. The photometry 
and redshift measurements for these sources are summarized in 
Table~\ref{table:src1}.  

\section{Data and Analysis}
\label{sec:daa}

The Giant Metrewave Radio Telescope (GMRT) L-band ($1.4$ GHz) receiver was used 
to observe the H~{\sc i} emission in this field. The observations were carried 
out on July 11 and 12, 2009 (GMRT observing cycle 16). Total observation 
duration was about 20 hours (10 hours on each day) with on-source time of 
about 12 hours. VLA calibrator source $1120+143$ ($\sim 5^\circ$ away from the 
target source) with $S_{1.4 {\rm GHz}} = 2.4$ Jy was used for phase 
calibration. This calibrator was observed for 7 minutes for every 45 minutes 
observation of the target field centred at galaxy-1. A standard flux calibrator (3C147 or 3C286) 
was observed for about 10-15 minutes in every 2 hours during the observation. 
The flux calibrator scans were also used for bandpass calibration. A total 
baseband bandwidth of $8.0$~MHz divided into $128$ frequency channels centered 
at $1328.35$~MHz, which corresponds to redshifted H~{\sc i} 21~cm frequency at 
$z = 0.0693$, was used for the observation. This gives a velocity resolution 
of $\sim 15$~km~s$^{-1}$ per channel and a total velocity coverage of $\sim 
1800$ km\ s$^{-1}$. Data analysis was carried out using standard {\small AIPS}. 
After flagging out bad data, the flux density scale, instrumental phase and 
frequency response were calibrated. The calibrated visibility data from the 
two days were combined and used to make an image cube and check for H~{\sc i} 
emission. The line-free channels were used to make continuum images at various 
resolution by weighting down different long baseline data points with Gaussian 
function using {\small UVTAPER} and {\small UVRANGE} in {\small IMAGR}. The 
high resolution continuum map has an rms noise of $\sim 142 
\mu$Jy beam$^{-1}$ for a synthesized beam size of $3.49\arcsec \times 
2.78\arcsec$. No continuum emission was detected in either high or low 
resolution images at the optical positions coinciding with that of these 
galaxies. The continuum emission from other sources in the field was subtracted 
from the data cube in the visibility domain. This continuum subtracted data 
cube, which contains signal only from the line emission, was imaged at 
different resolution to get the H~{\sc i} emission spectra and the H~{\sc i} maps. 
The H~{\sc i} mass detection limit for this observation ($\sim 4\times10^9 {\rm M}_\odot$) is better than that of the ALFALFA survey ($\sim 1.2\times10^{10} {\rm M}_\odot$) and comparable to that of the AGES for galaxies at a 
distance of $\sim 300$ Mpc \citep{co08}.

\section{Results}
\label{sec:rsl}

\begin{figure}
\begin{center}
\includegraphics[scale=0.30, angle=-90.0]{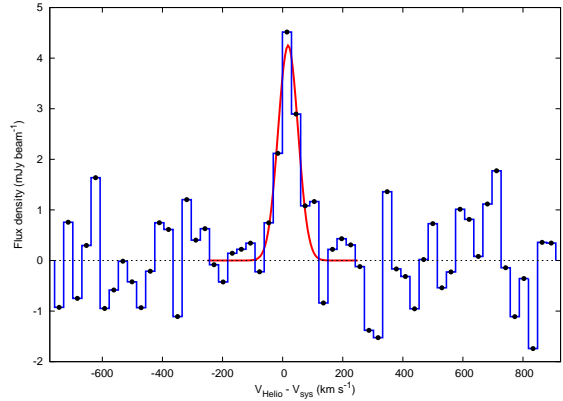}
\caption{\label{fig:gp2} H~{\sc i} emission spectrum for galaxy-1 (blue) and the best fit Gaussian function to the observed spectra (red). The linewidth ($\sigma$) is only $33$ km~s$^{-1}$ (full width at zero intensity $\sim 200$ km~s$^{-1}$).}
\end{center}
\end{figure}

\begin{figure}
\begin{center}
\includegraphics[trim = 0mm 17mm 0mm 17mm, clip, scale=0.40, angle=0.0]{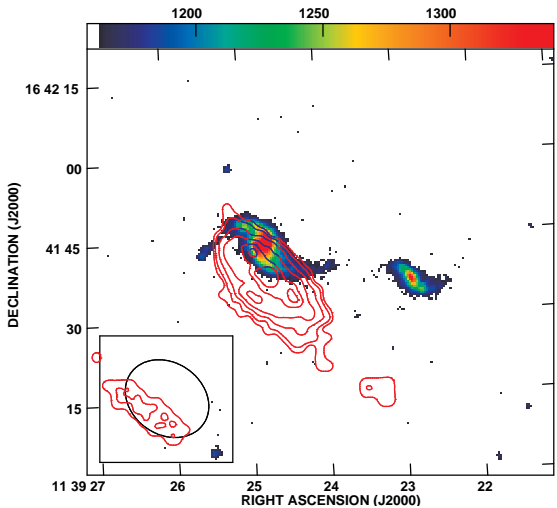}
\caption{\label{fig:gp3} SDSS DR6 $r$-band image of the galaxy pair J113924.74$+$164144.0 and J113922.85$+$164136.3 (galaxy-1 and galaxy-2) overlaid with the integrated H~{\sc i} emission in contours. For the H~{\sc i} map, the synthesized beam size is $16.6\arcsec \times 13.5\arcsec$ and the contour levels are for column densities of (3, 7, 15, 25, 35, 40) times $3.95\times10^{19}$ cm$^{-2}$. The low level H~{\sc i} contours are likely to be artifacts.}
\end{center}
\end{figure}

\begin{figure}
\begin{center}
\includegraphics[trim = 0mm 17mm 0mm 17mm, clip, scale=0.40, angle=0.0]{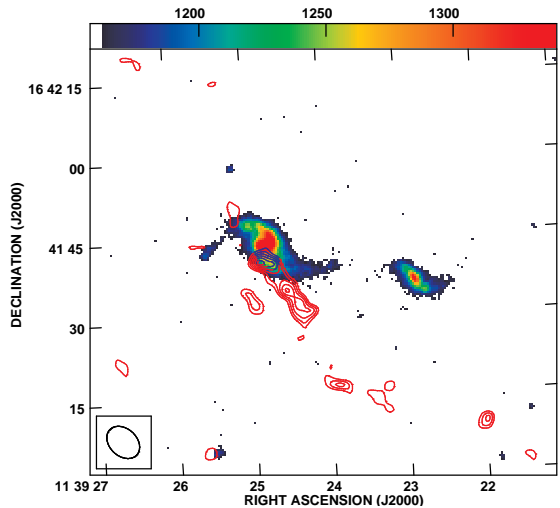}
\caption{\label{fig:gp4} SDSS DR6 $r$-band image of the galaxy pair J113924.74$+$164144.0 and J113922.85$+$164136.3 (galaxy-1 and galaxy-2) overlaid with the integrated H~{\sc i} emission in contours. For the H~{\sc i} map, the synthesized beam size is $6.9\arcsec \times 5.1\arcsec$ and the contour levels are for column densities of (5, 7, 9, 11, 13, 15) times $2.51\times10^{20}$ cm$^{-2}$. The low level H~{\sc i} contours are likely to be artifacts.}
\end{center}
\end{figure}

H~{\sc i} emission from galaxy-1 is detected at 
$> 6\sigma$ in the spectral channels with velocity close to the systemic 
velocity of the source. Figure~\ref{fig:gp2} shows the H~{\sc i} emission spectrum with a spectral resolution of $\sim 30$~km~s$^{-1}$ for this 
galaxy and the best fit Gaussian function to the data. The linewidth ($\sigma$) 
of the component is only $33$ km~s$^{-1}$ (full width at zero intensity $\sim 200$ km~s$^{-1}$). 
Luminosity distance $D_L$ of this galaxy for standard $\Lambda$CDM cosmology 
($H_0 = 73$ km s$^{-1}$, $\Omega_m = 0.27$, $\Omega_\Lambda =  0.73$) quoted 
in the NASA/IPAC Extragalactic Database (NED) is $305$ Mpc. Hence, the total 
H~{\sc i} mass for the galaxy is found to be 
\begin{equation}
M_{HI} = 2.36 \times 10^5~ D_L^2\int S_vdv ~~{\rm M}_\odot = 7.7\pm0.8 \times 10^9 {\rm M}_\odot
\end{equation}
where the distance to the galaxy $D_L = 305$ Mpc, $S_v$ is in Jy and the 
velocity interval is in km~s$^{-1}$. No H~{\sc i} emission is detected from 
the smaller spiral galaxy (galaxy-2) for which the $3\sigma$ upper 
limit of H~{\sc i} mass assuming a similar distance and linewidth is 
$2.4\times10^9 {\rm M}_\odot$. Since galaxy-2 is about factor of two 
smaller in size than galaxy-1, for a similar H~{\sc i} surface density, the total mass is 
expected to be $\sim1.9\times10^9 {\rm M}_\odot$, consistent with the derived 
upper limit. Figure~\ref{fig:gp3} and \ref{fig:gp4} show the SDSS 
DR6 $r$-band image of the field overlaid with the integrated 
H~{\sc i} emission contours for synthesized beam size of $16.6\arcsec 
\times 13.5\arcsec$ and $6.9\arcsec \times 5.1\arcsec$ respectively. Both 
these H~{\sc i} maps show the emission to be extended and significantly offset 
from the optical position of the galaxy. 

\begin{figure}
\begin{center}
\includegraphics[scale=0.30, angle=-90.0]{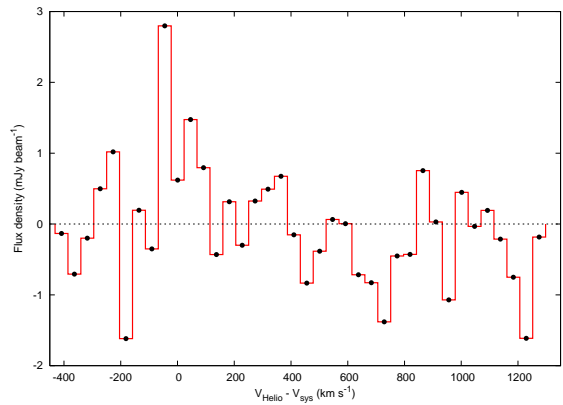}
\caption{\label{fig:gp5} H~{\sc i} spectrum for galaxy-7 with a resolution of $\sim 45$~km~s$^{-1}$.}
\end{center}
\end{figure}

\begin{figure}
\begin{center}
\includegraphics[trim = 0mm 17mm 0mm 17mm, clip, scale=0.40, angle=0.0]{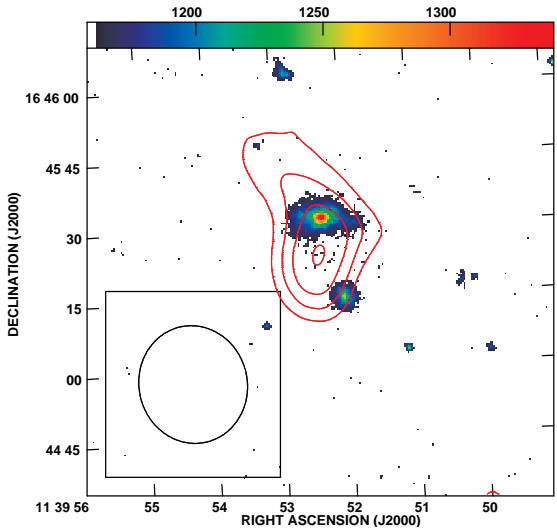}
\caption{\label{fig:gp6} SDSS DR6 $r$-band image of the galaxy pair J113952.31$+$164531.8 and J113952.02$+$164514.8 (galaxy-7 and galaxy-8) overlaid with the H~{\sc i} emission in contours for the channel at $V \approx 20350$ km~s$^{-1}$ and channel width $\sim 45$ km~s$^{-1}$. For the H~{\sc i} map, the synthesized beam size is $25.2\arcsec \times 23.0\arcsec$ and the contour levels are for flux densities of (2.5,3.0,3.4,3.8) times $400~\mu$Jy~beam$^{-1}$.}
\end{center}
\end{figure}

Figure~\ref{fig:gp5} shows the H~{\sc i} emission spectrum for galaxy-7 
with a spectral resolution of $\sim 
45$~km~s$^{-1}$ and a $> 4\sigma$ peak flux density of $\sim 2.8$ mJy. This 
marginally significant emission feature at the expected systemic velocity of 
the galaxy coinciding with the optical position makes it most likely to be 
H~{\sc i} emission from this galaxy. The emission profile shows weak 
indication of asymmetry and the total estimated H~{\sc i} mass is $5.3\pm1.3 
\times 10^9 {\rm M}_\odot$ for a luminosity distance of $299$ Mpc quoted in 
NED. Figure~\ref{fig:gp6} shows the SDSS DR6 $r$-band image of the field 
overlaid with the H~{\sc i} emission contours for a 
synthesized beam size of $25.2\arcsec \times 23.0\arcsec$. Apart from the 
disturbed optical morphology, for this galaxy also there is some hint that the 
H~{\sc i} emission is extended and offset from the optical position. But this 
offset is smaller than the beam size and hence needs to be confirmed. 
This low surface brightness emission is resolved in the high resolution map, 
making it difficult to draw any definitive conclusion on interaction between 
this pair of galaxies. 

One intriguing thing here is the striking similarity in the morphology of the 
central object of our study (galaxy-1) and that of the other 
spiral galaxy (galaxy-2) in the field. The small spiral galaxy-2 looks like a 
scaled down image of galaxy-1 with similar 
tidal-tail-like extended features. The total H~{\sc i} mass estimated for 
galaxy-1 from the GMRT observations is 7.7$\times$ 10$^9$ 
M$_{\odot}$. As has been mentioned earlier, the H~{\sc i} mass is mostly 
concentrated on galaxy-1 and seems to have an offset with 
respect to the optical position which can be due to a tidal interaction 
between the two neighbours. From the optical image, the galaxy diameter quoted 
in NED is $\sim 22\arcsec$. Assuming the entire H~{\sc i} is only from 
galaxy-1, the H~{\sc i} surface density of this galaxy turns out 
to be $\log(M_{\rm HI}/d_l^2) = 6.86$ M$_{\odot}$ kpc$^{-2}$, where $d_l$ is 
the linear diameter of the galaxy. The quantity $\log(M_{\rm HI}/d_l^2)$ is 
known to be a good diagnostic of the H~{\sc i} content of galaxies. Comparing 
the derived H~{\sc i} surface density for this galaxy with the expected 
surface density for spiral galaxies of similar size \citep{hg84}, we find no 
H~{\sc i} deficiency in this particular case. Thus even if we detect a 
reasonably disturbed and displaced H~{\sc i} disk in the main galaxy of the 
system possibly due to an ongoing interaction, we do not see any signs of gas 
loss from the system. This conclusion is based on the assumption that the 
entire H~{\sc i} detected belongs to galaxy-1 and there is no contribution 
from the other galaxy. But, its external origin can not be ruled out 
completely. A possible explanation of the disturbed and displaced H~{\sc i} 
disk of galaxy-1, presumably used to be a nicely rotating spiral, is tidal 
interaction with galaxy-2. Based on the absence of $m = 2$ symmetry and the 
fact that its center being too off from the optical center, it is less likely 
to be caused only through tidal interaction. Gas accretion from satellites 
(potentially including galaxy-2) may also contribute to the origin of this 
displaced H~{\sc i} gas.

These two systems at $z_{sp} = 0.0680$ and $0.0693$ are most likely to be 
distinct and evolving independently. However, the other possibility is that 
they have been involved in an interaction. The projected separation is $\sim 
600$ kpc which is a reasonable size for a moderate group. Assuming the true 
distance to be few times $600$ kpc and typical velocity to be few hundred 
km~s$^{-1}$, the corresponding timescale is $\sim 6$ Gyr. So, if there was any 
interaction, the close approach would have happened $\sim 6$ Gyr ago. It is 
also worth mentioning that since the photometric redshift measurements often 
have large errors, it is possible that all these galaxies almost at the same 
redshift are in a loose group and interacting with each other. 

\section{Conclusions}
\label{sec:con}

We have reported H~{\sc i} observation of the field with an interesting 
galaxy J113924.74$+$164144.0 at $z = 0.0693$, and two other neighbouring 
galaxies, a spiral galaxy J113922.85$+$164136.3 which has a strikingly similar 
`tidal' morphology, and a faint galaxy J113923.58$+$164129.9. Narrow H~{\sc i} 
emission is detected from J113924.74$+$164144.0, but 
J113922.85$+$164136.3 shows no detectable emission. The total H~{\sc i} mass 
detected in J113924.74$+$164144.0 is about $7.7\times10^9$ M$_\odot$. The 
H~{\sc i} emission from the galaxy is found to be extended and offset from the 
optical position of the galaxy. This, along with the long, tidal-tail-like 
extended optical features indicate possibility of interaction with the 
neighbouring spiral galaxy. About $7\arcmin$ away from these sources there is 
another galaxy J113952.31$+$164531.8 at $z = 0.0680$ with a close neighbour 
J113952.02$+$164514.8. There is a possible H~{\sc i} detection of the 
galaxy J113952.31$+$164531.8 with an estimated H~{\sc i} mass of about 
$5.3\times10^9$ M$_\odot$. The scenario, that all these galaxies are almost at 
the same redshift and interacting with each other, can not be ruled out due to 
possible large uncertainties of photometric redshifts. Deep optical 
spectroscopic measurements are required to ascertain the redshifts of all the 
galaxies.

\section*{Acknowledgements}

We thank the staff of the GMRT who have made these observations possible. GMRT 
is run by the National Centre for Radio Astrophysics of the Tata Institute of 
Fundamental Research. We are grateful to the anonymous referee for useful 
comments and for prompting us into substantially improving this paper. NR is 
also grateful to Jacqueline H. van Gorkom and Eric W. Greisen for useful 
discussions. NR is a Jansky Fellow of the National Radio Astronomy Observatory 
(NRAO). The NRAO is a facility of the National Science Foundation operated 
under cooperative agreement by Associated Universities, Inc.


\label{lastpage}

\end{document}